\documentstyle[12pt,epsf,rotate]{article}
\begin{document}

\begin{center}
{\bf\Large  A Burst of Electromagnetic Radiation \\
            from a Collapsing Magnetized Star\\}
\ \\
G. V. Lipunova
\\
\ \\
Sternberg Astronomical Institute, Universitetskii pr. 13,\\
Moscow, 119899 Russia\\
\ \\
\ \\
 Astronomy Letters V. 23 (1997) P.104\\
\end{center}
\bigskip
\bigskip

\begin{abstract}
The pattern of variations in the intensity of
magnetodipole losses is studied with the relativistic effect of
magnetic-field dissipation during collapse into a black hole
taken into account. A burst-type solution can be obtained both
for a direct collapse and for the formation of a
rapidly-rotating, self-gravitating object --- a spinar --- using a
simple model. Analytical dependences on radius describing an
electromagnetic burst are derived. The time dependence of the
burst shape for an infinitely distant observer and the maximum
energy of relativistic particles accelerated by an electric
field are numerically calculated. The objects under
consideration are of particular interest because particles in
their vicinity can be accelerated up to the Planck energies.
Possible astrophysical applications to the theory of active
galactic nuclei (AGNs) and QSOs are briefly discussed. It is
shown for the first time that a spinar can be produced by a
merger of neutron stars; this possibility is considered in and
without connection with the formation of gamma-ray bursts.
\end{abstract}

\section{Introduction}
The collapse of a star with substantial angular momentum can be
accompanied by the formation of a quasi-static object whose
equilibrium is maintained by centrifugal forces. Such objects
(spinars or magnetoids) were first considered in connection with
attempts to explain the activity of QSOs and AGNs in terms of
the model of supermassive ($\sim 10^6 \div  10^9 M_{\odot}$)
stars (Hoyle and
Fowler 1963; Ozernoi 1966; Morrison 1969; Ozernoi and Usov 1973;
Lipunov 1987). Ostriker (1970) and Lipunov (1983) suggested the
existence of low-mass spinars with nearly solar masses.

It is well known that neither the gas pressure nor the magnetic
pressure can inhibit the once started collapse, because they
grow no faster than the density of gravitational energy of the
collapsing star. Only fairly rapid rotation can hinder the
catastrophic contraction; in this regard, a spinar, as we call
it for definiteness below, is just such a rapidly-rotating
object. In fact, this implies that we consider a low-entropy
object. A consideration of a "hot" spinar may be the subject of
a separate study.

The lifetime of "cool" spinars is determined by the
characteristic time of angular-momentum loss. If a spinar
possesses a strong magnetic field and there is no substantial
mass outflow, then magnetodipole radiation may prove to be a
major dissipative mechanism; in this case, the evolutionary
parameters of the spinar can be estimated most simply (Woltjer
1971; see below).

The subsequent observations of QSOs and AGNs over a wide
wavelength range, particularly in X rays and gamma rays, have
provided convincing evidence for the model of a supermassive
accreting black hole, and the idea that spinars exist in nature
has lost its luster.

However, the question of whether spinars exist (if only
temporarily) is far from being ultimately resolved, at least for
several reasons, some of which have been elucidated most
recently.

First, supermassive rapidly-rotating magnetized stars may be the
evolutionary precursors of supermassive black holes in the
nuclei of galaxies and QSOs. This scenario is considered in
connection with the formation of supermassive stars at an epoch
that preceded the formation of galaxies in the Universe as a
result of growth of the initial perturbations after
recombination (see, e.g., Loeb and Rasio 1994). Second, Ozernoi
and Lipunov (1996) has recently pointed out that conditions for
the generation of limiting magnetic and electric fields
characteristic of the "cosmic supercollider" (Kardashev 1995),
in which particles can accelerate to superhigh Planck energies,
can be naturally created in the vicinity of strongly magnetized
supermassive spinars; this is of fundamental importance for
modern physics.

In addition, it has been noted earlier (Lipunov 1983) that a
large excess of angular momentum arises during the induced
collapse of an accreting white dwarf, which is capable of
halting the direct collapse with the subsequent temporary
formation of a spinar of nearly solar mass.

Finally, an excess momentum can be produced by orbital motion
during a merger of neutron stars, which are currently considered
as a possible source of gamma-ray bursts; this is noted in this
paper for the first time. The latter possibility is particularly
attractive in that the presence of strong magnetic fields seems
natural and that the ejected mass may be completely absent due
to the high binding energy of matter in neutron stars. After the
merger, the energy losses from the newly formed object --- a
spinar --- will be governed by the mechanism of magnetodipole
radiation. The intensity of magnetodipole losses, which is
proportional to the fourth power of the rotation frequency, for
a solar-mass spinar with $\nu=1000$~Hz and the same magnetic
flux as that of the Crab pulsar will then be $10^6$~Crab. Thus, a
spinar can be observed from a distance of $60$~Mpc, given that
currently available radio telescopes can detect sources that are
weaker than the Crab pulsar by a factor of $1000$. According to
the current scenarios for the evolution of binary stars, the
frequency of mergers of neutron stars is of the order of one
merger per $10^4$ years (Lipunov et al. 1987; Tutukov and
Yungelson 1992; van den Heuvel 1994). Thus, we can observe the
formation of a spinar several times per year.

The aim of this work is to study magnetodipole radiation of a
spinar with the relativistic effect of magnetic-field
dissipation during the formation of a black hole taken into
account. We will consider magnetodipole radiation as a major
factor that governs the evolution of a spinar. Our primary
concern will be the evolution of magnetodipole luminosity of a
spinar, or, more precisely, the evolution of the luminosity
calculated using the magnetodipole formula. In this case, it
should be remembered that the experience of studying radio
pulsars shows that we are actually dealing with the complicated
physical process of relativistic-particle acceleration with the
ensuing pronounced nonthermal radiation over a wide wavelength
range; at the same time, however, what is particularly
remarkable is that the energy losses are well described
precisely by the magnetodipole formula as before.

As a result (using a relatively simple physical model), we will
obtain a complex curve for the evolution of a spinar's
luminosity in the form of a nonmonotonic burst that describes
the salient features of a real astrophysical process. We will
also consider the limiting case of direct collapse of a star
with a magnetic field that is dipolar at a distance and finally
discuss some astrophysical applications of our model.

\section{A classical burst}
The discovery of QSOs and AGNs in 1963 marked the beginning of
an active study of black holes and their precursors. It was
noted in several papers that the gravitational collapse of
magnetized stars produces a burst of electromagnetic radiation.
Novikov (1964) considered the direct collapse of a star with
magnetodipole moment. He found the intensity of magnetodipole
radiation to take the form of a monotonically increasing
(formally to infinity) burst,

$$     L = \Phi ^2{c\over 6R_g} \left({R_g\over R}\right)^4,  $$
where

$$           \Phi ={\mu \over R}=const.   $$
The second equality gives the law of conservation of the
magnetic flux through the star, $\mu$ is the magnetodipole
moment of a spherically-homogeneous star. The stellar surface is
involved in a free fall and is described by the law of motion of
Newtonian form
$$        \ddot R = - {GM\over R^2}. $$

Generally speaking, a burst will also occur during the collapse
of a spinar. Hoyle and Fowler (1963), Ozernoi (1966), and
Ozernoi and Usov (1973) considered supermassive spinars as the
precursors of black holes in AGNs.

A decrease of a spinar's angular momentum as a result of some
dissipative processes leads to an increase of the angular
rotation velocity $\omega$ and, hence, to an increase of the
kinetic rotational energy. In other words, spinars, like any
gravitationally-bound object, have negative thermal capacity.

Let us write out the Newtonian system of basic equations that
describe a spinar with a magnetic field characterized by the
magnetodipole moment  $\mu$ (Bisnovatyi-Kogan and Blinnikov 1972;
Woltjer 1971; Lipunov 1987):
$$  {\omega^2}R={GM\over R^2},\eqno(1) $$

$$     {\mu \over R}=const,\eqno(2)  $$

$$     {dI\omega \over dt}=-{2\over 3c^3} \mu^2 \omega^3,\eqno(3)   $$

$$     I=AMR^2.\eqno(4)  $$

Equation (1) is the condition of equilibrium for a point at the
equator. As was noted above, neglect of the contribution of
pressure is a simplification that can be justified by the fact
that only the centrifugal force can stop the catastrophic
contraction during the collapse. Equation (2) describes the law
of conservation of the magnetic flux through the spinar and
gives the pattern of variations in the magnetodipole moment,
$$    \mu = \mu_o{R\over R_o}. \eqno(5)   $$
$I$ is the moment of inertia, $A$ is close to unity, and equation
(3) expresses the dipole nature of the losses of angular
momentum $I\omega$. The solution of this system of equations is
given by

$$  R=R_o\left(1-{t\over t_k}\right)^{1/3},\qquad \omega=\omega_o
\left({t_k\over t_k-t}\right)^{1/2},
$$
where $\omega_o$ and $R_o$ are the initial rotation frequency and
radius of the spinar, respectively, and
{$t_k={Ac^3\over G} {1\over B_o^2R_o}=
4.05\cdot 10^8 B^{-2}_{12} R^{-1}_6 $}~s;
$Bo$ is the initial polar magnetic field, $B_{12}=B_o/10^{12}$~G,
and $R_6=R_o/10^6$~cm. Evidently, the rotation frequency of
the spinar becomes infinite in a finite time $t=t_k$; at the
same time, the intensity of magnetodipole radiation also
increases infinitely,

$$   L={2\over 3c^3} \mu^2 \omega^4={2\over 3c^3} \mu_o^2 \omega_o^4
\left({t_k\over t_k-t}\right)^{4/3},
$$
where $\mu_o=B_o R_o^3/2$ is the initial magnetodipole moment
of the spinar. Note that the collapse time of a spinar increases
considerably compared to that of a nonrotating object, for which
it is determined by the time of a free fall of the stellar
boundary:

$$  R=R_o\left(1-{t\over t_f}\right)^{2/3},\qquad
 t_f={2R_o^{3/2}\over {3\sqrt{2GM}}}\,
\approx 4.08\cdot 10^{-5}R_6^{3/2}M_{\ast}^{-1/2}~{s,} $$
where $M_{\ast}$ is the mass in units of $M_{\odot}$.

How does the energy of a spinar's magnetic field changes during
the collapse? Let the following condition be fulfilled at some
instant in time:

$$   {{B_o^2}\over 8\pi} {{4\pi}\over 3} R^3=k{GM^2\over R},         \eqno(6)
$$
where $k=U_{magn}/U_{grav}$. Evidently, this condition is
equivalent to the condition of conservation of the magnetic
flux, and, consequently, it is fulfilled throughout the entire
collapse, $k=const\approx 6.2\cdot 10^{-13} M_{\ast}^{-2} B_{12}^2 R_6^4$.

\section{Magnetic field in a spherically--symmetric case in the general
theory of relativity}
As was shown above, the magnetodipole moment of the magnetized
objects under consideration (a nonrotating star and a spinar),
which linearly varies with radius, becomes zero at $R = 0$.
Outside the scope of the classical treatment, we must take into
account the fact that a black hole possesses no magnetic field
that could be detected by an external observer: "a black hole
has no hairs". Thus, the dependence of $\mu$ on the radius of a
magnetized star must reflect the fact that $\mu = 0$ at $R =R_g$.

The behavior of a magnetic field near the gravitational radius
can be elucidated by solving Maxwell's equations in the general
theory of relativity. A magnetostatic solution for a
spherically-symmetric object, i.e., in the Schwarzschild metric,
was found by Ginzburg and Ozernoi (1964).

In the Schwarzschild metric, Maxwell's equations in the general
theory of relativity take the form

$$ {\partial F_{\theta \varphi}\over \partial r}+
{\partial F_{\varphi \theta}\over \partial \theta} = 0,
\qquad  {\partial\over \partial r}\left(r^2\sin\theta F^{\varphi r}\right)
+ {\partial\over \partial \theta}\left(r^2\sin\theta F^{\varphi \theta}\right)
=0,
$$
given the following conditions: (a) we seek an axially-symmetric
solution, and, hence, the electromagnetic-field tensor $F_{ik}$
does not depend on $\varphi$; (b) $F_{\theta r}=0$ or $B_\varphi=0$,
i.e., we seek a magnetic-dipole field that is uniform at
infinity; and (c) the current density is taken to be zero
throughout, with the exception of the stellar surface.

Recall the following relations:
$$B_r=\sqrt{{\,\sl g}_{\theta \theta}}\sqrt{{\,\sl g}_{\varphi \varphi}}
F^{\theta \varphi},\qquad
  B_\theta=\sqrt{{\,\sl g}_{\varphi \varphi}}\sqrt{{\,\sl g}_{rr}}
F^{\varphi r},\qquad
  B_\varphi=\sqrt{{\,\sl g}_{rr}}\sqrt{{\,\sl g}_{\theta \theta}}
F^{r \theta},$$
$$  {\,\sl g}_{\theta \theta}=r^2,\qquad {\,\sl g}_{\varphi \varphi}=
r^2\sin^2\theta,
        \qquad {\,\sl g}_{rr}=(1-r_g/r)^{-1},
$$
where ${\sl g}_{ik}$ is the metric tensor in the Schwarzschild metric.
Considering the properties of the solution noted above, one
assumes that
$$ B_r=2\cos\theta r^{-3} f(r)\mu  \eqno(7)   $$
and
$$ B_\theta=\sin\theta r^{-3} \psi(r)\mu, \eqno(8)  $$
where the functions $f(r)$ and $\psi(r)\,\to 1$ as $r\to \infty$. The
solution consists of two linearly independent parts (we will
consider only the radial component):

a uniform field $B_o$ inside a sphere, for which
$$ f(r)={B_o\over 2\mu}r^3,  \eqno(9)  $$
and an external part:
$$ f(r)=-3 \left(r\over {R_g}\right)^3 \left\{\ln\left(1-{R_g\over r}\right)+
{R_g\over r} + {1\over 2}\left(R_g\over r\right)^2 \right\} \equiv
-3\left(r\over {R_g}\right)^3 \xi(R_g/r).\eqno(10) $$
At infinity, we obtain the classical expression $ B_r =$ $2\mu/ r^3 $
for $\theta=\pi/2$.

Based on these expressions, we derive the dependence $\mu(R)$,
where $R$ is the current radius of the sphere under
consideration. Since the normal component of ${\bf B}$ does not change
at the intersection of the stellar surface, the polar field
outside will be equal to the uniform field inside. Let us write
this condition using (7) and (10) for $\theta=0$,
$$ B^{(in)}= 2\mu (-3) R_g^{-3} \xi(R_g/R). \eqno(11) $$
Assuming that the star is a fairly good conductor, we use the
law of conservation of the magnetic flux,
$$  2\pi \int{ B^{(in)} r dr} = const. $$
where $r$ is the Schwarzschild coordinate inside the star. We
obtain
$$    B^{(in)} R^2 = B^{(in)}_o R_o^2\, \eqno (12)   $$
Here, $R_o$ is some initial radius. From (11) and (12), we derive
$$   \mu(R) = -{1\over 6} R_g^3 \left(R_o\over R\right)^2 {B_o\over \xi(R_g/R)}.
\eqno (13) $$
or
$$   \mu(R) = \mu_o \left(R_o\over R\right)^2 {{\xi(R_g/R_o)}\over \xi(R_g/R)}.
\eqno (14) $$

It can be seen from formulas (7), (10), and (13) that the
magnetic moment decreases to zero as $R$ approaches $R_g$, while
the field is pressed to the star.

Thus, we derived the formula that describes magnetic-field
dissipation at a finite radius. Let us use it to calculate the
intensity of magnetodipole losses
$$  L = -{2\over 3} {{\ddot{\vec \mu}^2}\over {c^3}}, $$
and then consider two limiting cases:\\
(a) for a direct spherically-symmetric collapse, the change in
\hbox{${\vec \mu }$} is due to a decrease in the radius alone;\\
(b) for a spinar --- a rotating object --- ${\ddot{\vec \mu}}$ is determined by
rotation of the magnetic-dipole vector.

In both cases, we will solve the problem in a quasi-static
approximation, because we use the solution of a magnetostatic
problem. For this approach, we have a simple, qualitatively
adequate description of the decrease in the external magnetic
field during the collapse of a magnetized object. Note that in
the case of a spherically-symmetric collapse, an exact solution
exists for the field, according to which the magnitude of the
magnetic dipole decreases with time as $t^{-5}$ (Price 1972).

\section{Direct collapse}
In this case, we assume that the boundary of the star is
involved in a free fall,
$$
  {dR\over dt}=c \left(1-{R_g\over R}\right)\sqrt{R_g\over R}\, ,
 \eqno (15) $$
when $R_g/R_o\ll 1$, $t$ is the time of an infinitely distant
observer, because it is this observer who records the luminosity
$L$; at the same time,
$${{d^2\mu}\over dt^2} = {d^2\mu\over dR^2}\left({dR\over dt}\right)^2+
      {d\mu\over dR}{d^2 R\over dt^2}\, .\eqno(16)
$$
Using (15) and (16), we obtain
$$ L={2\over 3c^3} {{\mu^2(x) c^4}\over R_g^4} \eta^2(R_g/R) $$
or
$$ L=L_o \left({x\over x_o}\right)^4 {\xi^2(x_o)\over \xi^2(x)}
 {\eta^2(x)\over \eta^2(x_o)},
\eqno (17)$$
where $L_o$ is some luminosity at the initial time, at which
$x=x_o=R_g/R_o$. The expression for $ \eta(R_g/R)\equiv \eta(x) $
is given in the Appendix (formula (23)). Figure 1
shows that the radiation has the shape of a nonmonotonic burst.
The ratio $R/R_g$ at which the intensity reaches its maximum is
always the same in the approximation $R_o\gg R_g$ (this follows
from the fact that the variables $x$ and $x_o$ in expression (17)
are separated) and ${R_{max}/R_g}\approx 2.09$.

We can pass to the time dependence if the law of motion of the
stellar surface is known,
$$ t=t_o+ {R_g\over c} \left\{-{2\over 3y\sqrt{y}}-{2\over \sqrt{y}}
           +\ln{{1+\sqrt{y}}\over 1-\sqrt{y}} \right\}, $$
$$  t_o =-{R_g\over c} \left\{-{2\over 3y_o\sqrt{y_o}}-{2\over \sqrt{y_o}}
           +\ln{{1+\sqrt{y_o}}\over 1-\sqrt{y_o}} \right\},
           \qquad y_o = R_o/R_g
$$
and $ y_o\gg 1$. Let us rewrite $L$ with $k=U_{magn}/U_{grav}$,
$$ L={k\over 36}{c^5\over G}{x^4 \eta^2(x)\over \xi^2(x)},\qquad
   L_{max}=k\cdot 3.7\cdot 10^{59} \, \hbox{erg/s.}
$$

Note that the intensity of magnetodipole radiation for an
infinitely distant observer becomes an exact zero in an infinite
time, because the star contracts to the gravitational radius
infinitely long for this observer.

\section{A collapsing spinar}
The problem of magnetodipole radiation of a collapsing spinar in
the general theory of relativity differs markedly from the case
of a spherically-symmetric collapse considered in the preceding
section. The reason is that the spinar rotates, and, hence, the
Kerr metric should be used. However, we do not know how the
magnetodipole moment looks in the Kerr metric. It is not that
simple to solve Maxwell's equations in the Kerr metric.
Therefore, we use the solution of a magnetostatic problem in the
Schwarzschild metric, i.e., formula (13) for $\mu$ with the
qualification that this formula includes the main relativistic
effect: dissipation of a spinar's magnetic field for a distant
observer at a finite value of $R$.

So, let us write the system of approximate equations for a
spinar,
$$ {{dI\omega}\over d\tau}=-{2\over 3c^3}\mu^2(R) \omega^3,$$
$$ \mu(R)=\mu_o \left({R_o\over R}\right)^2 {\xi(R_g/R_o)\over \xi(R_g/R)},
\eqno (18)$$
$$ I= AMR^2 ,                                            $$
$$ \omega^2 = {MG\over R^3},                           $$
where $\tau$ is the time of a local observer that is at rest
with respect to a sphere of radius $R$. In the Schwarzschild
metric, stable circular orbits, for which Kepler's law has the
same form as in the Newtonian mechanics, exist up to $r=3R_g$.
In the Kerr metric, stable circular orbits can exist up to $r=R_g/2$;
in this case, Kepler's law in geometrical units takes
the form
$$ \Omega = {\sqrt{M}\over {r^{3/2}+a(R_g/2)^{3/2} } }.\eqno(19)  $$
where $a\simeq 1$ is the ratio of the total angular momentum of the
spinar to $M^2$ that is greater than $1$. Since expression (19)
does not differ much from Kepler's classical law, we will make
use of it.

The intensity of magnetodipole losses can be written as an
explicit function of the spinar's radius,
$$ L=L_o \left({R_o\over R}\right)^{10} {\xi^2(R_g/R_o)\over \xi^2(R_g/R)},
\eqno (20)$$
where $L_o$ is some luminosity at the initial time, at which $R=R_o$.

Thus, we obtained the curve with a distinctive burst-like shape
(Fig. 2a). Note that for any initial parameters of a collapsing
spinar, the ratio $R/R_g $ at which the intensity reaches its
maximum is the same (much as it was in the preceding section;
this follows from the fact that the variables $R/R_g$ and
$R_o/R_g $ in expression (20) are separated) and ${R_{max}/R_g}\approx 1.29$.

From the law of angular-momentum loss, we derive the law of
motion $\tau=\tau (R)$ by integration; an explicit, but
cumbersome form of this law is given in the Appendix (formula
(24)).

In order to determine the burst shape for an infinitely distant
observer, we must pass from the local time $\tau$ to the
Schwarzschild time $t$ of the infinitely distant observer using
the formula
$$   dt= \sqrt{1-R_g/R}\,d\tau. $$

For a \hbox{$M=3M_{\odot}$} spinar with a magnetic field that is standard for
pulsars, the pulse width at half maximum in time is
$\Delta t\approx 700$~years (Fig. 2b). Numerical simulations give
$$ \Delta t\approx 5\cdot 10^{-4} k^{-1}M_{\ast}~\hbox{ᥪ}.$$

With $k=U_{magn}/U_{grav}$, we can write
$$ L={k\over 144}{c^5\over G}{x^{10}\over \xi^2(x)},\qquad
   L_{max}=k\cdot 1.1\cdot 10^{57}\,  erg/s.
$$

Let us estimate the maximum energy to which particles can
accelerate in the magnetosphere of a spinar using relation (6),
$$   \varepsilon\sim e{v\over c}BR.\eqno(21a) $$

Assuming $ v=\omega R   $ and
$$ \omega={\sqrt{GM}\over R^{3/2}}, \qquad
   B={\sqrt{6kG}M\over R^2}  $$
we obtain
$$ \varepsilon\sim e\sqrt{6k}{G\over c} {M^{3/2}\over R^{3/2}}=
   e {\sqrt{3}c^2\over 2\sqrt{G}} x^{3/2} k^{1/2}= $$
$$  =9.05\cdot 10^{26}   x^{3/2} k^{1/2}\,  \hbox{eV},            \eqno(21b)
$$
where $x=R_g/R$ and $k=U_{magn}/U_{grav}$. However, the
corresponding magnetic field in the vicinity of low-mass spinars
for $R\sim R_g$ and $k$ approaching $1$ will be so strong that the
formation of elementary particles will begin, and, what is more,
the very existence of such strong fields is called into question
for such objects. At the same time, the maximum magnetic field
($k = 1$) for supermassive stars is considerably weaker,
$$
    B_{max}= {\sqrt{6} c^4\over 4G^{3/2}} {x^2\over M} \approx
               10^{19}{x^2\over M_{\ast}}~ \hbox{G}\, ,
$$
where $M_{\ast}$ is given in solar masses. For
\hbox{$M > 2.2\cdot 10^6 M_{\odot}$} \hbox{$B < 4\cdot 10^{13}$~G},
and the creation of electron--positron pairs may be
neglected. Thus, particles in the vicinity of supermassive
spinars can accelerate to energies that are typical of the early
Universe (Ozernoi and Lipunov 1996). A similar estimate was
first obtained by the magnetic field around a black hole by
Kardashev (1995).

At each fixed point $r$ in the Schwarzschild system of
coordinates, the radial component of ${\bf B}$ in the vicinity of a
spinar decreases as (see (7))
 $$ B_r=B_o \left({R_o\over R}\right)^2 {\xi(R_g/r)\over \xi(R_g/R)}
\cos{\theta}, $$
where $R$ is the radius of the spinar.

The accelerating potential at fixed distanced then varies during
the collapse in a burst-like manner (Fig. 3), $\theta=0$:
$$ V=B_r(r=2R){\omega R\over c} R =
{\sqrt{3}c^2\over 2\sqrt{G}} x^{3/2} k^{1/2}
 \left(r\over R\right)^2   {\xi (R_g/r)\over \xi(R_g/R)}\,.         \eqno(22)
$$

These estimates follow from the first, quasi-static
approximation to a real dynamical problem. A more rigorous
treatment must lead to a refinement of the results. Note also
that application of this approximation to a spinar seems more
justified that to a direct, spherically-symmetric collapse.

\section{Discussion}
Ozernoi and Usov (1973) showed that the intensity of
magnetodipole radiation versus radius during collapse has the
shape of a burst due to the angle between the magnetic and
rotation axes being reduced to zero.

However, the experience of theoretical and observational studies
of pulsars gained over the last twenty years suggests (Lipunov
1992) that the energetics of rapidly-rotating magnetized stars
depends only slightly on the orientation of the magnetic field
with respect to the rotation axis; it is mainly determined by
complex electrodynamical processes, such as the acceleration of
relativistic particles, the generation of electromagnetic and
Alfven waves, that are described in order of magnitude by the
magnetodipole formula (as, for example, in the Julian--Goldreich
model). Therefore, the derived intensities describe the energy
carried away by cosmic-ray particles and synchrotron radiation
over a wide spectral range (from radio to gamma rays) rather
than the low-frequency electromagnetic radiation.

Calculation of a particular shape of the spectrum of a
supermassive spinar presents a separate, complicated problem.
However, some distinctive features are clear even without a
detailed analysis. First of all, a magnetized spinar is the most
natural physical justification of the so-called phenomenological
model of an oblique rotator, which is widely used to interpret
the observed properties of QSOs and AGNs, especially those with
strong nonthermal radio emission and jets (see, e.g., Kardashev
1995).

Of course, the evolution of a spinar must inevitably lead to the
formation of a supermassive black hole, while this process can
be burst-like in nature (see above). Assuming that the
precursors of supermassive black holes possessed intrinsic
magnetic fields and considering that it takes an infinite time
for a black hole to approach its gravitational radius for an
external observer (a "frozen star"), any such black hole is
formally at some stage of evolution of its burst-type
magnetodipole radiation. Moreover, note that the burst duration
of $\sim 10^8$~years that is definitively related to the maximum
luminosity of a spinar, which is taken to be $\sim 10^{46}$~erg/s
(the standard luminosity of QSOs), corresponds to this
luminosity for given masses (Fig. 4). These values are now
considered to be typical lifetimes of QSOs. Note also that the
case of a spinar in a vacuum considered in this paper by no
means exhausts the whole variety of possible modes of
interaction of a magnetized spinar with the surrounding medium
(Lipunov 1987). The existence of at least six types of spinar
(an accretor, a propeller, an ejector, a superaccretor, a
superpropeller, and a superejector) holds a great variety of
astrophysical manifestations of supermassive magnetized objects.

Of special, fundamental interest is the possibility of particle
acceleration in the vicinity of supermassive spinars to the
Planck energies or, at least, to typical energies in the theory
of great unification ($10^{24}$~eV).

With reference to spinars formed by a merger of neutron stars,
in addition to a burst of gravitational radiation, pulsar-type
electromagnetic radiation caused by orbital motion is possible
even before the merger in a system of two neutron stars (Lipunov
and Panchenko 1996).

The nature of gamma-ray bursts is so far unclear. It is possible
that they result from mergers of neutron stars (Blinnikov @et
al. 1984; Paczynski 1991). The release of electromagnetic
energy considered above describes a burst with a duration from
fractions of a second to ten seconds, if the surface field of a
spinar with $M=3M_{\odot}$ reaches \hbox{$5\cdot 10^{16}\div 10^{17}$~G}
for $R_o=10$~km (Fig. 5).

Let us make some rough estimates for the formation of low-mass
spinars by a merger of neutron stars. How and why an
intermediate object that we call a spinar can form? Here, the
influence of two factors is possible:\\
(1) we do not know the exact value of the Oppenheimer--Volkov
limit. In other words, the pressure of nuclear matter may play a
major role in addition to rotation;\\
(2) it is possible that excess angular momentum that inhibits a
direct collapse will suffice for the formation of a spinar.

It should be noted that difficulties emerge when we consider the
magnitude of the transferred momentum, at least in the first
approximation. A black hole cannot have the Kerr parameter
\hbox{$a\equiv {K/M^2} > 1$}, where $K$ is the total angular momentum
(see, e.g., Shapiro and Teukolsky 1985). It is in this case that the
formation of a spinar is implied.

Let us first estimate the rotation parameter $a$ in the Newtonian
approximation. Calculate the total angular momentum of a system
of two identical neutron stars. The moment of inertia of two
identical bodies about their center of symmetry is
$$ I=2\{AmR^2+m(d/2)^2\},$$
where $m$ is the mass of each star, and $d$ is the separation
between the centers of mass of the stars. We assume that the
rotation of the objects around  a common center of mass and
around their axes is synchronous. The total angular momentum is
$K=I\Omega$, where $K=I\Omega$ is the frequency of revolution.
Kepler's law in geometrical units ($c=G=1$) takes the form
$$   \Omega = {\sqrt{2m}\over d^{3/2}}. $$
We shall consider a merger for $d=2R=2\kappa R_g$. We then have
$$ a= {K\over 4m^2} = (A+1)\sqrt{\kappa \over 8}.$$

(1) $A=0$ --- intrinsic rotation is ignored. In this case, the
condition for the formation of a spinar (the condition that a
black hole is not formed) takes the form
$$  \sqrt{\kappa \over 8}>1,\qquad \kappa=R/R_g $$
At the same time, $\kappa \sim 2\div 3$ for neutron stars.

(2) For homogeneous spheres, $A=2/5$ and $ a\approx \sqrt{\kappa / 4}$~.

(3) For flat disks, $A=1/2$ and $ a\approx \sqrt{\kappa / 3.5}$~.

Evidently, the orbital angular momentum in the Newtonian
approximation is not enough for a spinar to be formed. Let us
estimate the relativistic effects.

Consider a problem that is similar to the problem of reduced
mass in classical mechanics. We then have the motion of a body
of mass $\mu=m/2$ in the effective potential of a gravitating
body of mass $M=2m$. In the last stable circular orbit in the
Schwarzschild metric, the corresponding angular momentum is
$l=2\sqrt{3}\mu M$ (Zel'dovich and Novikov 1971). Then, we have
$K=2\sqrt{3} m^2$ and $a=\sqrt{3}/2$.

However, a more rigorous treatment of the problem using the
general theory of relativity with allowance for a more or less
real equation of state of matter is needed to establish what
angular momentum the two neutron stars had before the merger.

\bigskip\bigskip\bigskip\bigskip

\section*{Acknowledgments}

I wish to thank Prof. V.M. Lipunov for the formulation of the problem.
I am also grateful to the referees for helpful critical remarks and
to V. Astakhov for the translation.

\section*{Appendix}

\begin{enumerate}
\item In formula (17), which gives the law of motion of the surface
in a free collapse,
$$ \eta(x)=x(1-x)(6x(1+x)-x^2(1-3x))+{{2x^9}\over \xi^2(x)}+
    {{3x^6(1-x)}\over \xi(x)},\eqno(23)$$
where
$$ \xi(x) =  \ln(1-x)+ x + {1\over 2}x^2,\qquad
 x={R_g\over R}.  $$
\item Integrating (18) gives the equation of motion for the surface
of a spinar in explicit form:
$$  -{1\over 9}(x^{-9}+1)ln^2(1-x)+{2\over 9}\left\{\sum_{k=1}^8{1\over kx^k}+
  ln\left({1-x}\over x\right)\right\}ln(1-x)-$$
$$  -{2\over 9}\sum_{k=1}^8{1\over k}\left\{\sum_{m=1}^k{1\over mx^m} +
   ln\left({{1-x}\over {x}}\right)\right\}-
   {2\over 9}\sum_1^{\infty}{(1-x)^k\over k^2}-$$
$$   -\left({x^{-8}\over 4}+{x^{-7}\over 7}\right)ln(1-x)-
    {x^{-7}\over 7}-{x^{-6}\over 6}-{x^{-5}\over 20}+$$
 $$  +{{11}\over 28} ln\left({{1-x}\over x}\right)
   +{1\over 4}\sum_{k=1}^7{1\over kx^k}+
    {1\over 7}\sum_{k=1}^6{1\over kx^k}=$$
$$   ={1\over 216}{{c^3 R_o^4 B_o^2}\over A G^2 M^3}(\tau+\tau_o),
\quad\eqno(24)$$
where $x=R_g/R$, $\tau=0$ at $x=R_g/R_o$.
\end{enumerate}

\section*{References}
\noindent
Bisnovatyi--Kogan G.S., Blinnikov S.I. {\it Astrophys.
and Space Sci.} 1972. V. 19. P. 119.
\noindent
Blinnikov S.I., Novikov I.D., Perevodchikova T.V. ¨ Polnarev A.G.
{\it Soviet Astron. Lett.} 1984. V. 10. P. 177.\\
\noindent
Woltjer I. {\it Pont. Acad. Sci. Nucl. Galaxies.} 1971. V. 35. P. 477.\\
\noindent
Ginzburg V.L., Ozernoy L.M. {\it Zhurn. Eksper. Teor. Fiziki}.
1964. V. 47. P. 1030.\\
\noindent
Zeldovich Ya.B., Novikov I.D. The theory of gravitation
and star evolution. Œoscow: Nauka, 1971.\\
\noindent
Kardashev N.S. {\it Monthly Notices Roy. Astron. Soc.}
1995. V. 276. P. 515.\\
\noindent
Lipunov V.M. {\it Astrophys. and Space Sci.} 1983. V. 97. P. 121.\\
\noindent
Lipunov V.M. The astrophysics of neutron stars. 1992. Springer--Verlag,
Heidelberg.\\
\noindent
Lipunov V.M. {\it Astrophys. and Space Sci.} 1987. V. 132 P. 1.\\
\noindent
Lipunov V.M., Postnov K.A., Prokhorov M.E. {\it Astron. and Astrophys.}
1987. V. 176. L1.\\
\noindent
Lipunov V.M., Panchenko I.E. {\it Astron. and Astrophys.} 1996. V. 312. P. 937. \\ 
\noindent
Loeb A. and Rasio F.A. {\it Astrophys.J.} 1994.
V. 432. P. 52.\\
\noindent
Morrison P. {\it Astrophys.J.} 1969. V. 157. P. L73\\
\noindent
Novikov I.D. {\it Sov. Astron. Zhur.} 1964. V. 41. P. 290.\\
\noindent
Ozernoy L.M. {\it Sov. Astron. Zhur.} 1966. V. 10. P. 241.\\
\noindent
Ozernoy L.M., Usov V.V. {\it Astrophys. and Space Sci.}
 1973. V. 25. P. 149.\\
\noindent
Ozernoy L.M., Lipunov V.M. {\it Bull. Amer. Phys. Soc.}
1996. V. 41. No. 2. P. 927.\\
\noindent
Ostriker J.P. {\it Proc. 11th Conf. Cosmic Rays.}
Budapest: Acta Phys. Acad. Sci., 1970. V. 29. P. 69.\\
\noindent
Paczynski B. {\it Acta.Astron.} 1991. V. 41. P. 257\\
\noindent
Price R.H. {\it Physical Review.} 1972. D5. Num.10.
P. 2419.\\
\noindent
Tutukov A.V., Yungelson L.R.) {\it Monthly Notices Roy.
Astron. Soc.} 1992. V.260. P. 675.
\noindent
van den Heuvel E.P.J. Interacting
Binaries. Eds Shore S.N., Livio M., van den Heuvel E.P.J.
Berlin: Springer, 1994. P. 263-474.\\
\noindent
Hoyle F., Fowler W.A. {\it Monthly
Notices Roy. Astron. Soc.} 1963. V. 125. P. 169.\\
\noindent
Shapiro S.L., Teukolsku S.A. Black Holes, White Dwarfs,
and Neutron Stras. Moscow: Mir, 1985\\

\clearpage

\section*{Figure captions}
\begin{itemize}

\item[Fig. 1.] The power of magnetodipole losses versus radius for a
spherically-symmetric collapse (formula (17)).
\bigskip

\item[Fig. 2.] (a) The power of magnetodipole losses versus radius for
a relativistic spinar (formula (20)); the power reaches its
maximum at $R/R_g=1.29$. (b) The power of magnetodipole losses
versus time for a relativistic spinar (the magnetic flux is
$\Phi=const=10^{12}\hbox{~G}\cdot\pi 10^{12}~cm^2$;
$k\approx 7\cdot 10^{-14}$.

\bigskip

\item[Fig. 3.] The accelerating electric potential (formula (22)) in
the vicinity of a limiting spinar versus radius.
\bigskip

\item[Fig. 4.] Evolution of the luminosity of a supermassive
relativistic spinar. For $R=1.5\cdot 10^{14}~cm$, the surface magnetic
field is $B\approx 1.8\cdot 10^4$~G. The width at half maximum in time is
$\approx 1.6\cdot 10^8$~years.

\bigskip

\item[Fig. 5.] The power of magnetodipole losses versus time for a
relativistic spinar. The magnetic flux is
$\Phi=const=5\cdot 10^{16}\hbox{~G}\,\cdot\pi 10^{12}~cm^2$;
$k\approx 2\cdot 10^{-4}$. The burst width at
half maximum is $8.89$~s.

\end{itemize}

\clearpage
\begin{figure}
\epsfysize=10cm
\centerline{\epsfbox{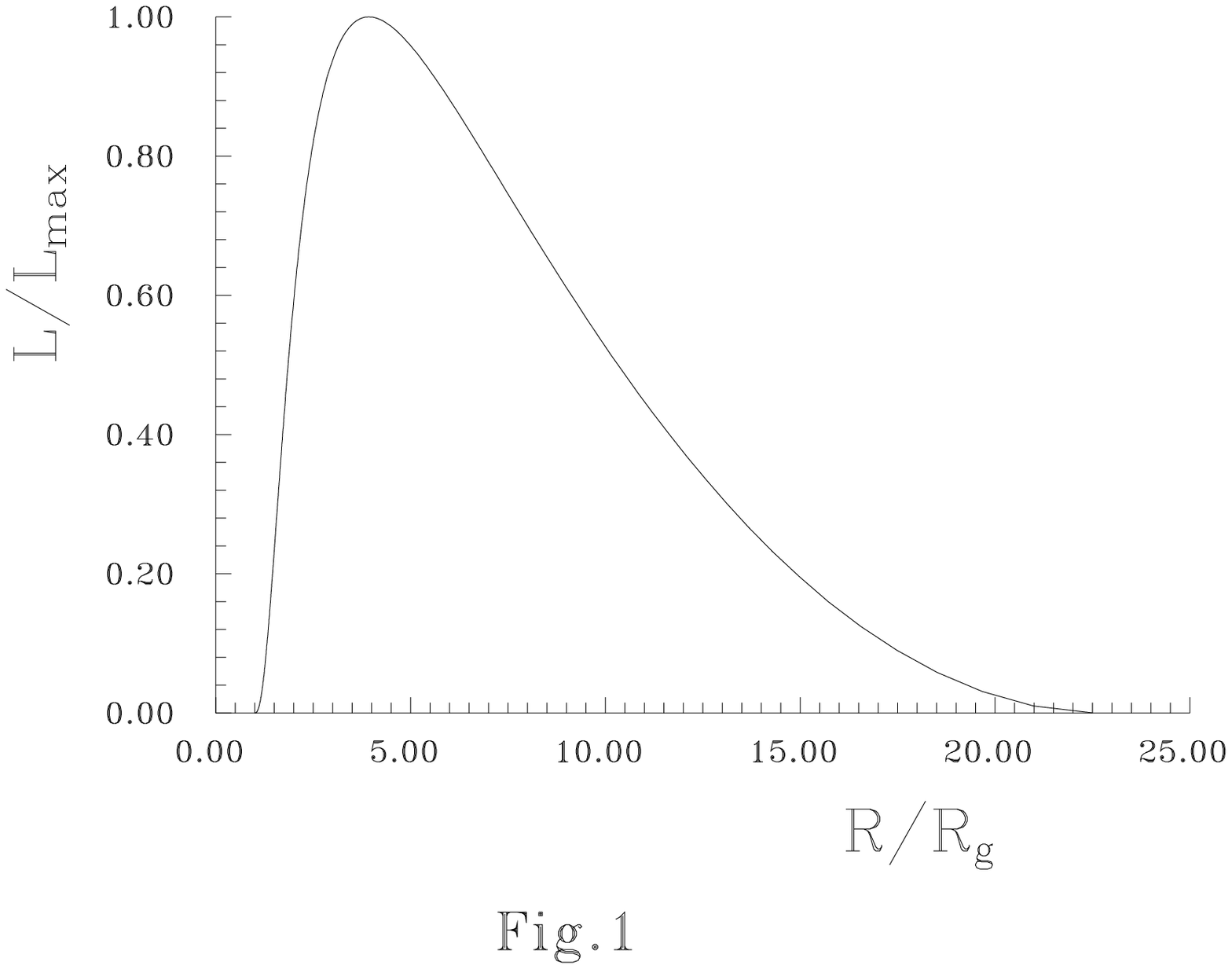}}
\end{figure}

\clearpage

\begin{figure}
\epsfysize=10cm
\centerline{\epsfbox{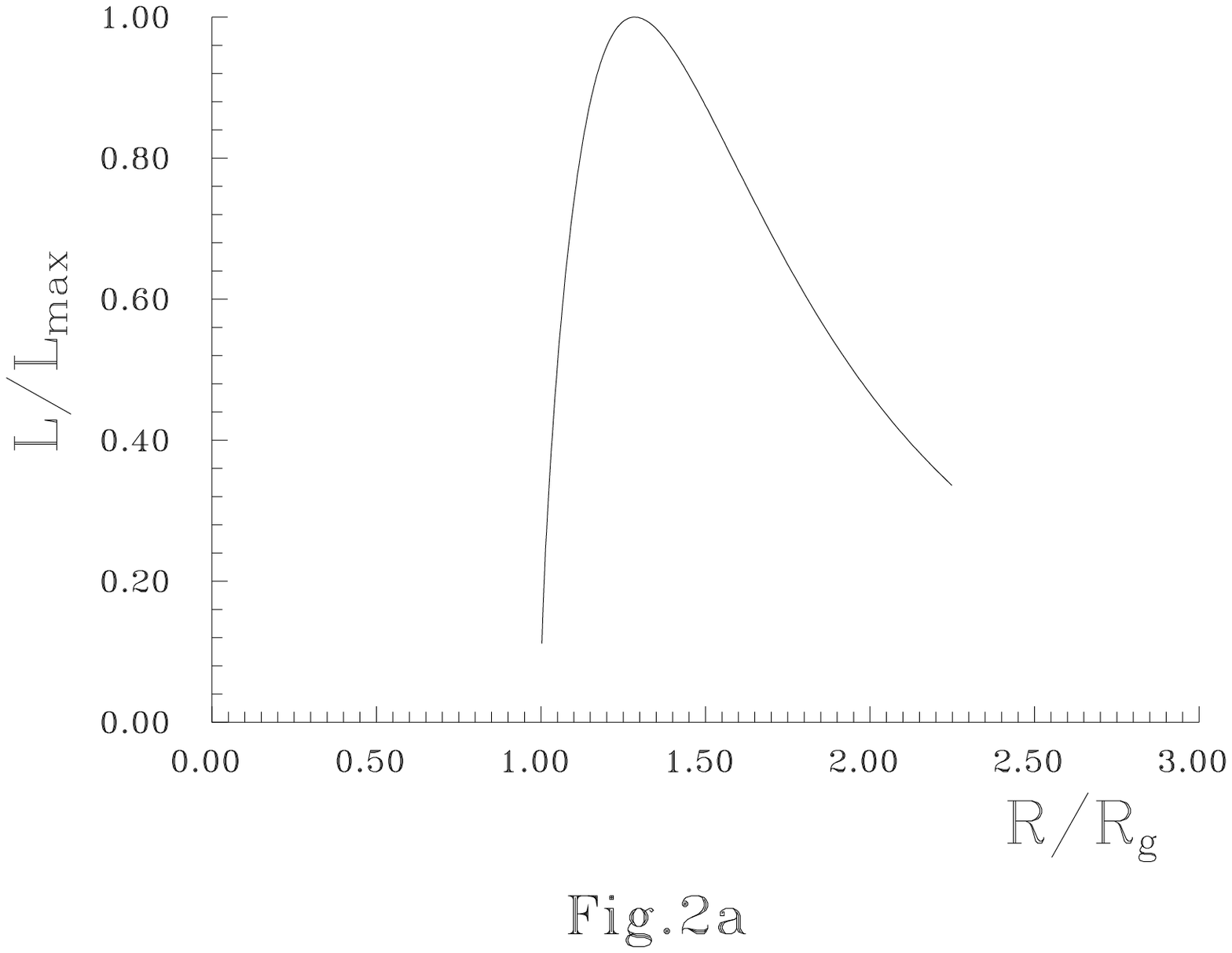}}
\end{figure}

\clearpage

\begin{figure}
\epsfysize=10cm
\centerline{\epsfbox{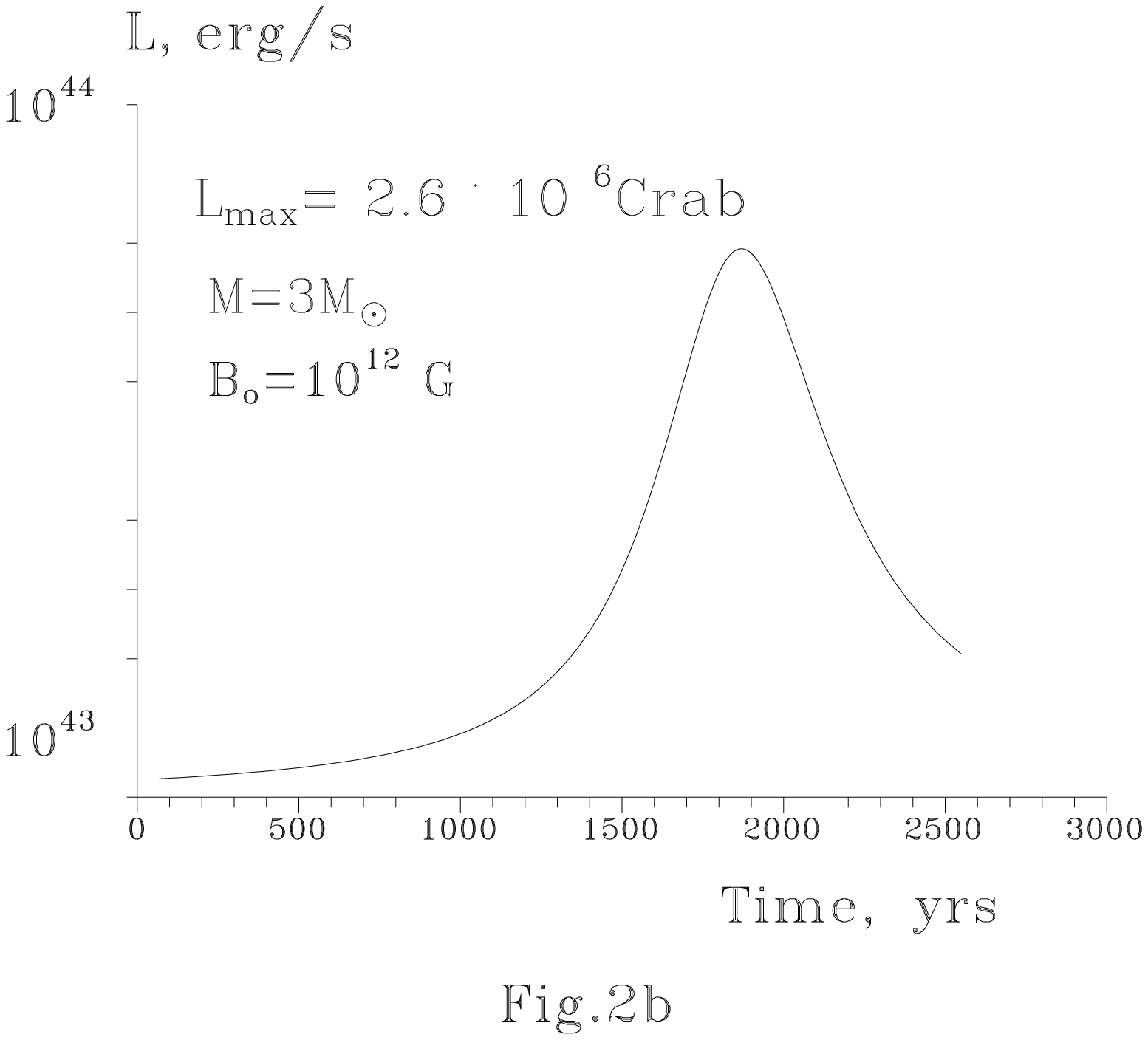}}
\end{figure}

\clearpage

\begin{figure}
\epsfysize=10cm
\centerline{\epsfbox{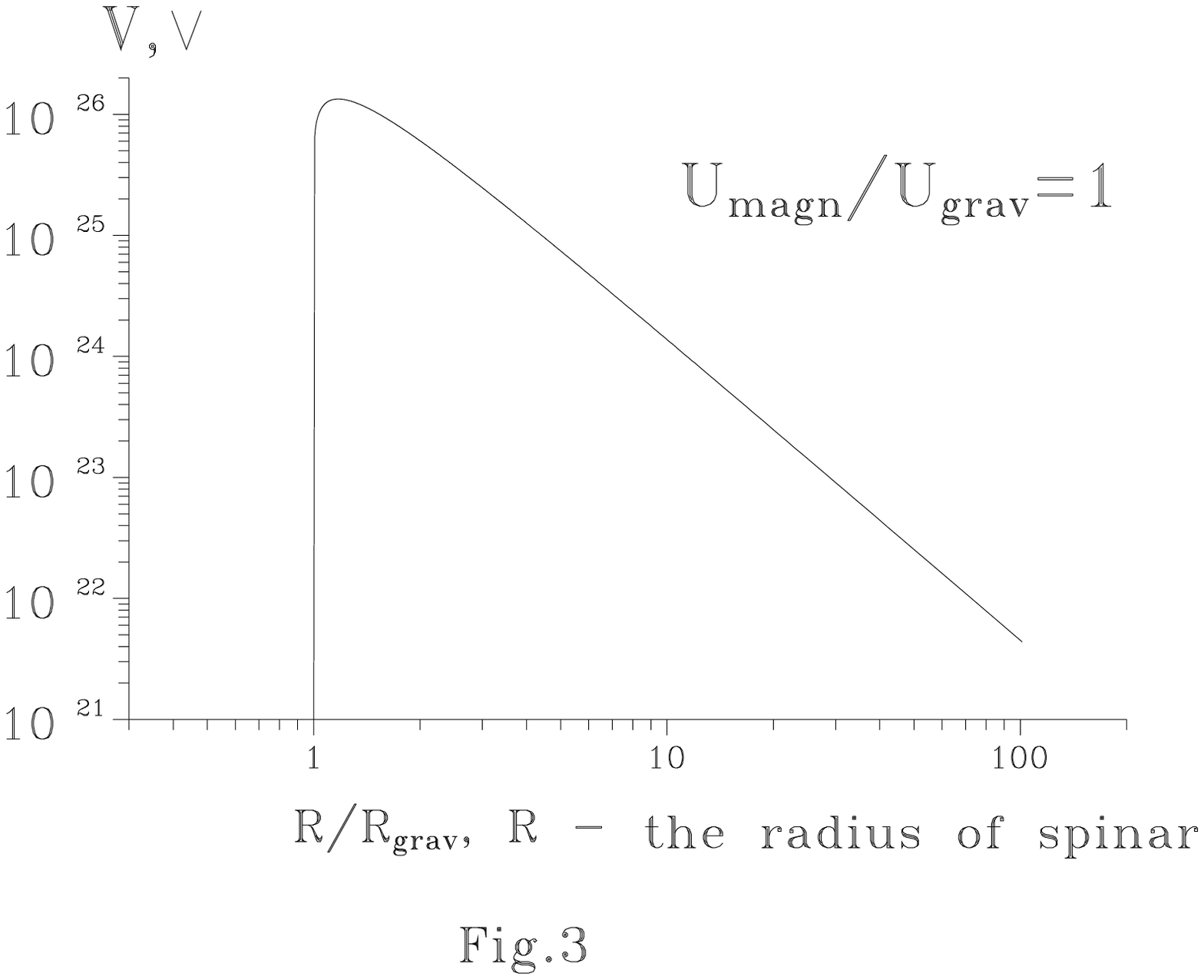}}
\end{figure}

\clearpage
\begin{figure}
\epsfysize=10cm
\centerline{\epsfbox{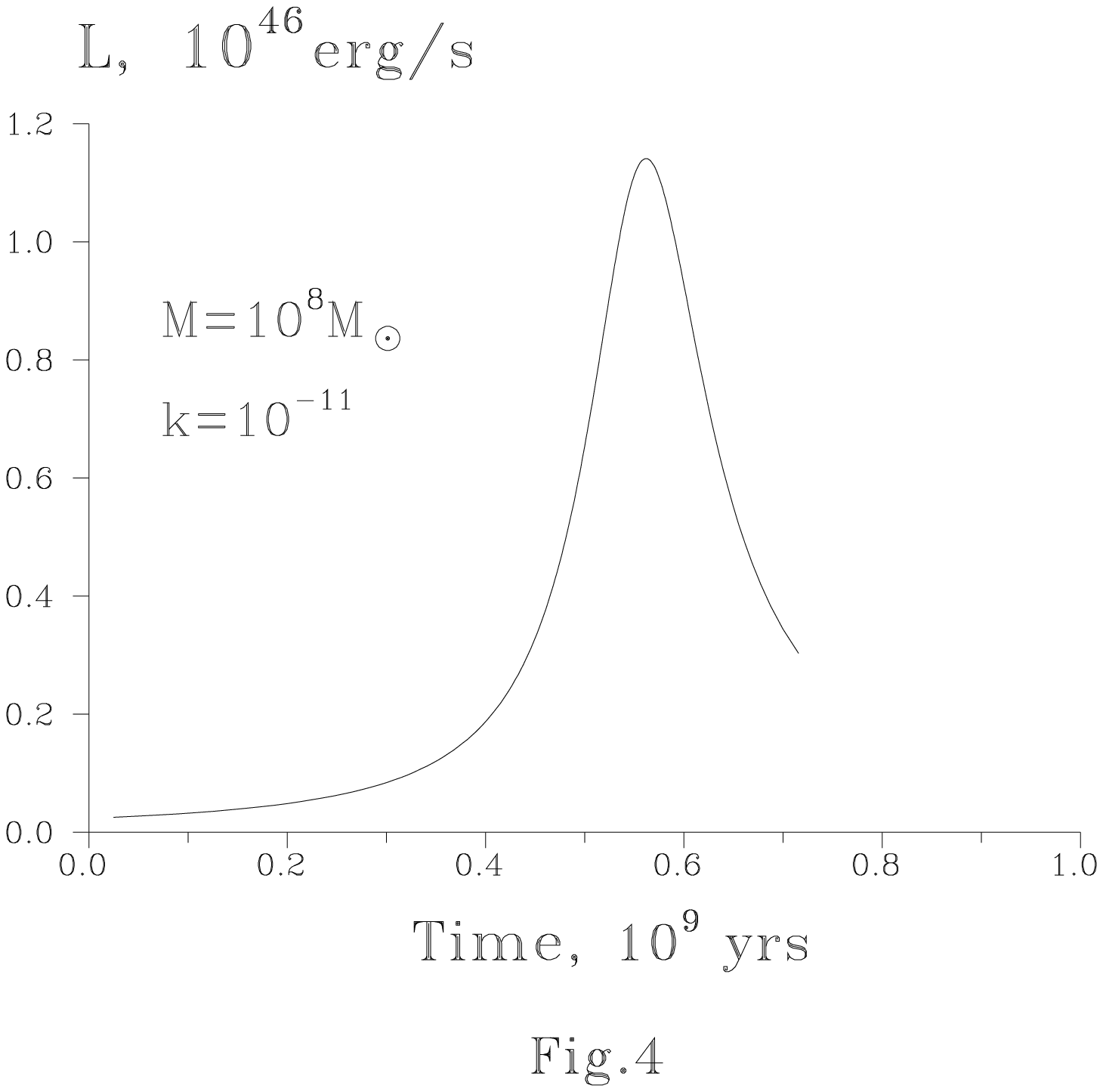}}
\end{figure}

\clearpage

\begin{figure}
\epsfysize=10cm
\centerline{\epsfbox{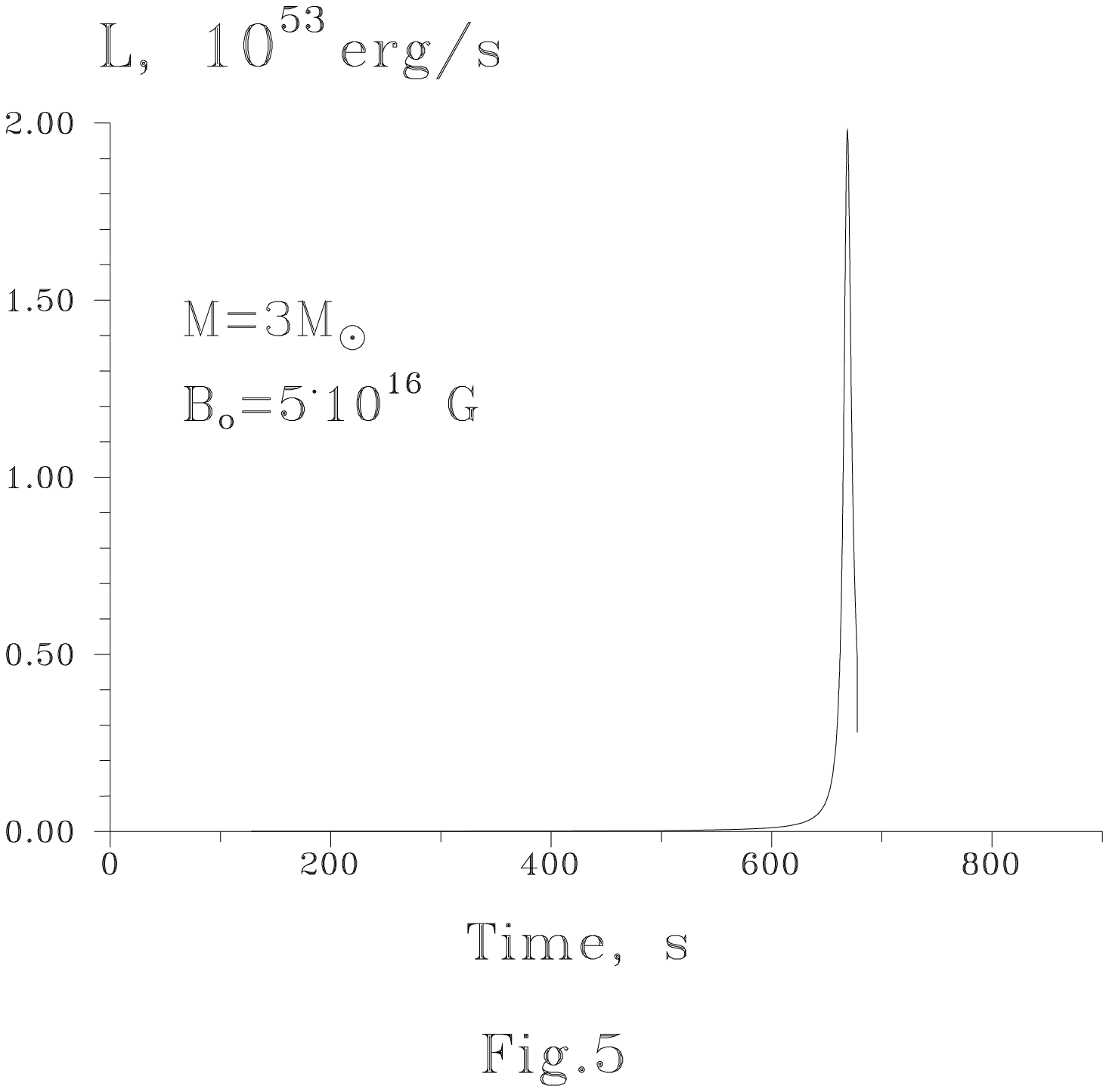}}
\end{figure}

\end{document}